\documentclass[conference, letterpaper]{IEEEtran}
\ifCLASSINFOpdf
\else
\fi
\ifCLASSINFOpdf
   \usepackage[pdftex]{graphicx}
\else
\fi

\usepackage[cmex10]{amsmath}
\usepackage{color}
\usepackage{fancyhdr}
\usepackage[caption=false,font=footnotesize]{subfig}
\usepackage{graphicx}
\usepackage{float}
\usepackage{booktabs}
\usepackage{listings}
\usepackage{array}
\usepackage{amsmath}
\usepackage{xcolor}
\usepackage{caption}

\renewcommand{\thispagestyle}[2]{}

\fancypagestyle{plain}{
        \fancyhead{}
        \fancyhead[C]{first page center header}
        \fancyfoot{}
        \fancyfoot[C]{first page center footer}
}
\usepackage{graphicx}   % for including figures
\usepackage{caption}  
\usepackage[font=footnotesize,labelfont=normalfont]{caption}
\begin{document}

%
% paper title
% can use linebreaks \\ within to get better formatting as desired
\title{State-of-the-Art in Software Security Visualization: A Systematic Review}

\author{\IEEEauthorblockN{Ishara Devendra\IEEEauthorrefmark{1}\IEEEauthorrefmark{2},
Chaman Wijesiriwardana\IEEEauthorrefmark{1},
Prasad Wimalaratne\IEEEauthorrefmark{3}}
\IEEEauthorblockA{\IEEEauthorrefmark{1}Department of Information Technology\\University of Moratuwa, Sri Lanka\\
 isharad@uom.lk, chaman@uom.lk}
\IEEEauthorblockA{\IEEEauthorrefmark{2}Faculty of Graduate Studies, University of Colombo, Sri Lanka}
\IEEEauthorblockA{\IEEEauthorrefmark{3}Department of Communication and Media Technologies\\University of Colombo School of Computing\\
Colombo 07, Sri Lanka\\
spw@ucsc.cmb.ac.lk}}

% use for special paper notices
%\IEEEspecialpapernotice{(Invited Paper)}

% make the title area
\maketitle

\begin{abstract}
Software security visualization is an interdisciplinary field that combines the technical complexity of cybersecurity, including threat intelligence and compliance monitoring, with visual analytics, transforming complex security data into easily digestible visual formats. As software systems get more complex and the threat landscape evolves, traditional text-based and numerical methods for analyzing and interpreting security concerns become increasingly ineffective. The purpose of this paper is to systematically review existing research and create a comprehensive taxonomy of software security visualization techniques through literature, categorizing these techniques into four types: graph-based, notation-based, matrix-based, and metaphor-based visualization. This systematic review explores over 60 recent key research papers in software security visualization, highlighting its key issues, recent advancements, and prospective future research directions.  From the comprehensive analysis, the two main areas were distinctly highlighted as extensive software development visualization, focusing on advanced methods for depicting software architecture: operational security visualization and cybersecurity visualization.  The findings highlight the necessity for innovative visualization techniques that adapt to the evolving security landscape, with practical implications for enhancing threat detection, improving security response strategies, and guiding future research.
\end{abstract}

\begin{IEEEkeywords}
Security Visualization; Vulnerability Analysis; Threat Intelligence; Compliance Monitoring
\end{IEEEkeywords}

% For peer review papers, you can put extra information on the cover
% page as needed:
% \ifCLASSOPTIONpeerreview
% \begin{center} \bfseries EDICS Category: 3-BBND \end{center}
% \fi
%
% For peerreview papers, this IEEEtran command inserts a page break and
% creates the second title. It will be ignored for other modes.
\IEEEpeerreviewmaketitle

\section{Introduction}

The increasing use of software has introduced significant security challenges, prompting the industry to develop best practices to protect applications that handle sensitive information. The critical need for robust software security is evident, as vulnerabilities can lead to severe consequences, including data breaches, unauthorized control, and service disruptions, thereby compromising security policies \cite{cui2022empirical, braz2022less, abeyratne2020security}. Although considerable research has been conducted to prevent these vulnerabilities and automate their detection, the effectiveness of these efforts remains limited due to the growing complexity of software systems \cite{senanayake2023android, wijesiriwardana2017detection}.

Software security visualization has emerged as a promising approach to address these challenges by providing intuitive and interactive visual representations of security data \cite{mortara2021visualization}. These techniques enhance the comprehension of security vulnerabilities, facilitate threat detection, and support the maintenance of secure codebases \cite{cobilean2023review}. However, despite its potential, there is a lack of comprehensive studies that systematically analyze the effectiveness of different software security visualization techniques and identify the specific areas where these techniques can be most impactful \cite{siavvas2023security}. 

This review aims to fill this gap by conducting a systematic analysis of existing research in the field of software security visualization. By examining key studies and identifying the strengths and weaknesses of current visualization techniques, this paper highlights the critical areas where further research is needed and suggests directions for future development.

The paper is organized as follows: Section II discusses the research methodologies, including the study selection criteria. Section III presents an analysis of the findings, organized into comprehensive software development visualization, operational security visualization, and cybersecurity visualization. Section IV explores how software security visualization techniques are applied in real-world contexts. Section V discusses the key findings, including the categorization of visualization methods and their impact. Section VI concludes by emphasizing the importance of visualization in addressing cybersecurity challenges and supporting effective threat analysis.

\section{Methodology}

This section outlines the systematic approach used to review relevant studies, ensuring a comprehensive and unbiased selection of sources. It outlines the process for identifying, evaluating, and selecting studies based on predefined criteria to ensure the reliability and validity of the review.

\subsection{Search Strategy}

To ensure a comprehensive literature review, a systematic search strategy was employed following the PRISMA guidelines \cite{page2021prisma}. Databases such as IEEE Xplore, ACM Digital Library, SpringerLink, and ScienceDirect were searched for articles published over the last 5 years (2019–2024). The search terms used included combinations of ``software security visualization,'' ``cybersecurity visualization tools,'' ``security data visualization,'' and related keywords. The search strategy was designed to capture all relevant studies in the field of software security visualization \cite{okoli2015guide, kitchenham2013systematic}.

\subsection{Eligibility Criteria}

Inclusion criteria for studies were rigorously defined as follows:

\begin{itemize}
    \item Published in peer-reviewed journals or conference proceedings \cite{page2021prisma}.
    \item Focus on visualization techniques specifically used in software security \cite{okoli2015guide}.
    \item Provision of empirical results concerning the effectiveness of visualization tools in cybersecurity.
\end{itemize}

Studies were excluded if they met any of the following criteria:

\begin{itemize}
    \item Non-English language publications.
    \item Opinion pieces, editorials, or non-empirical articles.
    \item Studies not directly related to the field of software security visualization \cite{kitchenham2013systematic,snyder2019literature}.
\end{itemize}

\subsection{Study Selection}

The initial search yielded a total of 120 records. A detailed multi-step selection process was then employed:

\begin{itemize}
\item Duplicate Removal: Automated and manual processes were used to identify and remove duplicate records, resulting in 90 unique studies \cite{page2021prisma}.

\item Title and Abstract Screening: The titles and abstracts of these studies were independently reviewed by two researchers to assess relevance. This step eliminated studies that did not align with the predefined inclusion criteria, resulting in 70 studies for full-text review \cite{okoli2015guide}.

\item Full-Text Review: The remaining articles were subjected to a detailed full-text review to assess their adherence to the eligibility criteria. This involved critically assessing the study’s methodology, focus, and contributions to the field.  In cases where the relevance was unclear, a third reviewer was consulted to reach a consensus \cite{brereton2007lessons}.

\item Final Selection: After the full-text review, 61 studies were deemed suitable for inclusion in the systematic review. The entire selection process was meticulously documented, and a PRISMA flow diagram was created to represent the screening and inclusion process visually \cite{kitchenham2007guidelines}.
\end{itemize}

\subsection {Data Extraction}

Data were extracted from each study by two independent reviewers using a standardized form \cite{page2021prisma}. The extracted information included authors, year of publication, study objectives, visualization techniques used, main findings, and contributions to software security visualization. Any discrepancies between reviewers were resolved through discussion or consultation with a third reviewer \cite{okoli2015guide}.

\subsection {Quality Assessment}

The quality of the included studies was assessed using the Critical Appraisal Skills Programme (CASP) checklist. This assessment focused on the methodological rigor and credibility of the studies, evaluating aspects such as the clarity of data presentation, the appropriateness of the analytical methods, and the generalizability of the findings \cite{munn2018systematic}.

\subsection {Synthesis of Results}

The data were synthesized qualitatively, categorizing studies by their focus areas: comprehensive software development visualization, operational security visualization, cybersecurity visualization, and information security visualization. Trends, commonalities, and divergences across studies were analyzed to identify the state of the art and gaps in the current literature \cite{kitchenham2007guidelines}.

\section{Overview of Software Security Visualization}

A systematic literature review has demonstrated the state-of-the-art of software visualization, aiding developers in selecting suitable visualization tools.

This study observed that most software visualizations are displayed on computer screens, and several approaches that are based on the structure of the system are displayed in 3D environments with city metaphors \cite{chotisarn2020systematic}. However, despite the advancements in visualization techniques, gaps remain, particularly in the application of these tools across different domains and user levels.

A study has reviewed 54 publications on Cyber Situational Awareness (CSA) visualizations, emphasizing the need for more comprehensive visualizations that support all levels of CSA and cater to a broader range of stakeholders. They identified gaps in existing CSA visualizations, such as a lack of visualizations for managers and non-expert users, and a focus on threat information over impact information and response plans \cite{jiang2022systematic}. This gap highlights the need for more user-centric visualization tools that non-technical stakeholders can effectively utilize, a critical area that remains underexplored.

\begin{figure}[H]
 \centering
 \includegraphics[width=100mm]{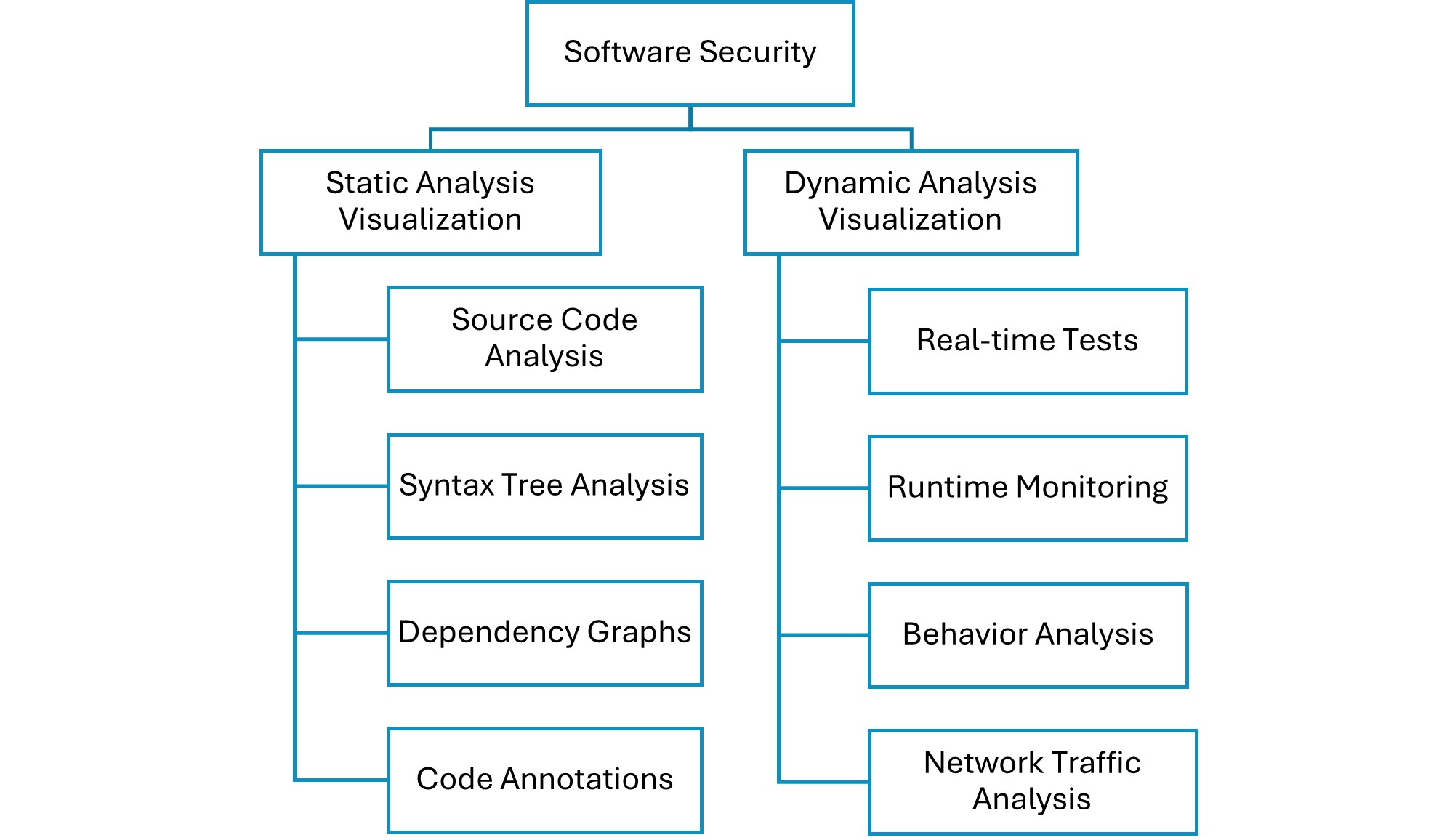}
 \caption{Dynamic and Static Analysis Visualization}
 \label{fig:Figure1}
\end{figure}

Another study has emphasized the need for standardized evaluation metrics for cybersecurity visualizations, highlighting methodological gaps and proposing potential research directions to establish a unified framework for evaluating the effectiveness of security visualizations \cite{sun2023cyber}. Despite these proposals, a consensus on standardized metrics is still lacking, leading to inconsistencies in how visualization tools are evaluated and compared across different studies. Figure \ref{fig:Figure1} categorizes software security visualization into two main types: Static Analysis Visualization and Dynamic Analysis Visualization.

\subsection{Static Analysis Visualization in Software Security}

Static analysis visualization is crucial for understanding and identifying static code-based vulnerabilities without executing the program. This approach improves the efficiency and effectiveness of security assessments by providing clear visual perspectives into complex code structures and potential security issues.

Mollerstrom and Roos highlighted the critical role of threat analysis, particularly in making potential threats more understandable and actionable. It focuses on visualizing data flow within a program to assist in manual code reviews, making it easier to identify security vulnerabilities \cite{mollerstrom2023engineering}. However, while visualization techniques are effective in static analysis, they cannot often detect runtime issues, limiting their applicability in dynamic environments.

Recent studies have further explored these methodologies in the context of container security, highlighting the use of static code analysis in identifying and mitigating vulnerabilities within software architectures \cite{wong2023containers}. Despite these advancements, the integration of static analysis with other security assessment tools remains a challenge, particularly in maintaining accuracy and scalability across different environments.

Similarly, Senanayake et al. reviewed methods for detecting vulnerabilities in Android source code, while emphasizing the role of visualization in presenting the identified vulnerabilities to the software practitioners \cite{senanayake2023android}. However, the effectiveness of these visualization techniques in large-scale software systems with complex dependencies has not been thoroughly examined, highlighting a gap in the current research.

Recent studies have highlighted the ability to use deep learning methodologies with visualization techniques to increase the efficiency of vulnerability detection \cite{zeng2020software,alaoui2022deep}. While deep learning enhances the accuracy of detection, its integration with existing visualization tools poses challenges, particularly in terms of interpretability and the potential for false positives.

Studies by Fedorchenko et al. and Piantadosi et al. discussed models and approaches for exploit analysis and smart contract security, respectively, highlighting the importance of visual representations in mitigating and understanding security risk \cite{fedorchenko2023analytical,piantadosi2023detecting}. These studies underscore the importance of context-aware visualizations that can adapt to different threat environments, a feature that is often missing in current tools.

\subsection{Dynamic Analysis Visualization in Software Security}

Dynamic Analysis Visualization plays a crucial role in identifying and understanding security vulnerabilities in real time when software executes. This approach enhances static code analysis by demonstrating runtime issues that might not be detected through static code analysis. Recent studies have highlighted the significance of understanding coding vulnerabilities among various languages and the utilization of vulnerability detection tools, respectively, both of which benefit from dynamic analysis in capturing real-time issues \cite{sakharkar2023systematic,sorensen2019literature}. However, these studies often lack a comparative analysis of how different tools detect complex, multi-stage attacks.

Alaoui and Nfaoui have noted that dynamic analysis enhanced with visualization can improve the detection of unknown threats, and the importance of using deep learning approaches in identifying web application attacks \cite{alaoui2022deep}. While this approach is promising, the challenges of integrating dynamic analysis with existing security frameworks and ensuring the scalability of these solutions are still underexplored. Similarly, Cerny et al. and Parker et al. demonstrate the importance of dynamic analysis to manage evolving system architecture and operational inefficiencies while discussing the reconstruction and visualization of microservice architecture and runtime anti-pattern identification, respectively \cite{cerny2022microservice,parker2023visualizing}. Some researchers have focused on detecting and visualizing complex vulnerabilities such as TOCTOU race conditions and multistage cyber-attacks, where dynamic analysis provides crucial insights into the temporal aspects of security threats \cite{raducu2022defense}. However, the effectiveness of these techniques in real-world scenarios, particularly in large-scale distributed systems, remains under-evaluated, representing another gap in the current research.

\subsection{Software Security Visualization Techniques/ Tools}

Software security visualization techniques and tools play a critical role in enhancing an understanding of security vulnerabilities in software systems. Recent studies have highlighted the significant importance of visualization techniques in understanding software architecture and evolution, categorizing them into graph-based, notation-based, matrix-based, and metaphor-based types \cite{cerny2022microservice}. While these categorizations are helpful, they often overlook the potential overlaps between different types of visualizations, which could lead to more integrated and effective tools. Some studies have focused on the requirements and evaluation of software visualization tools, highlighting the need for scalability, interactivity, and rigorous validation methods to ensure their effectiveness \cite{medeiros2023visualizations}. However, the lack of standardized evaluation metrics remains a significant barrier to comparing and improving these tools, as noted in the literature.

\begin{figure*}[ht]
 \centering
\includegraphics[width=140mm]{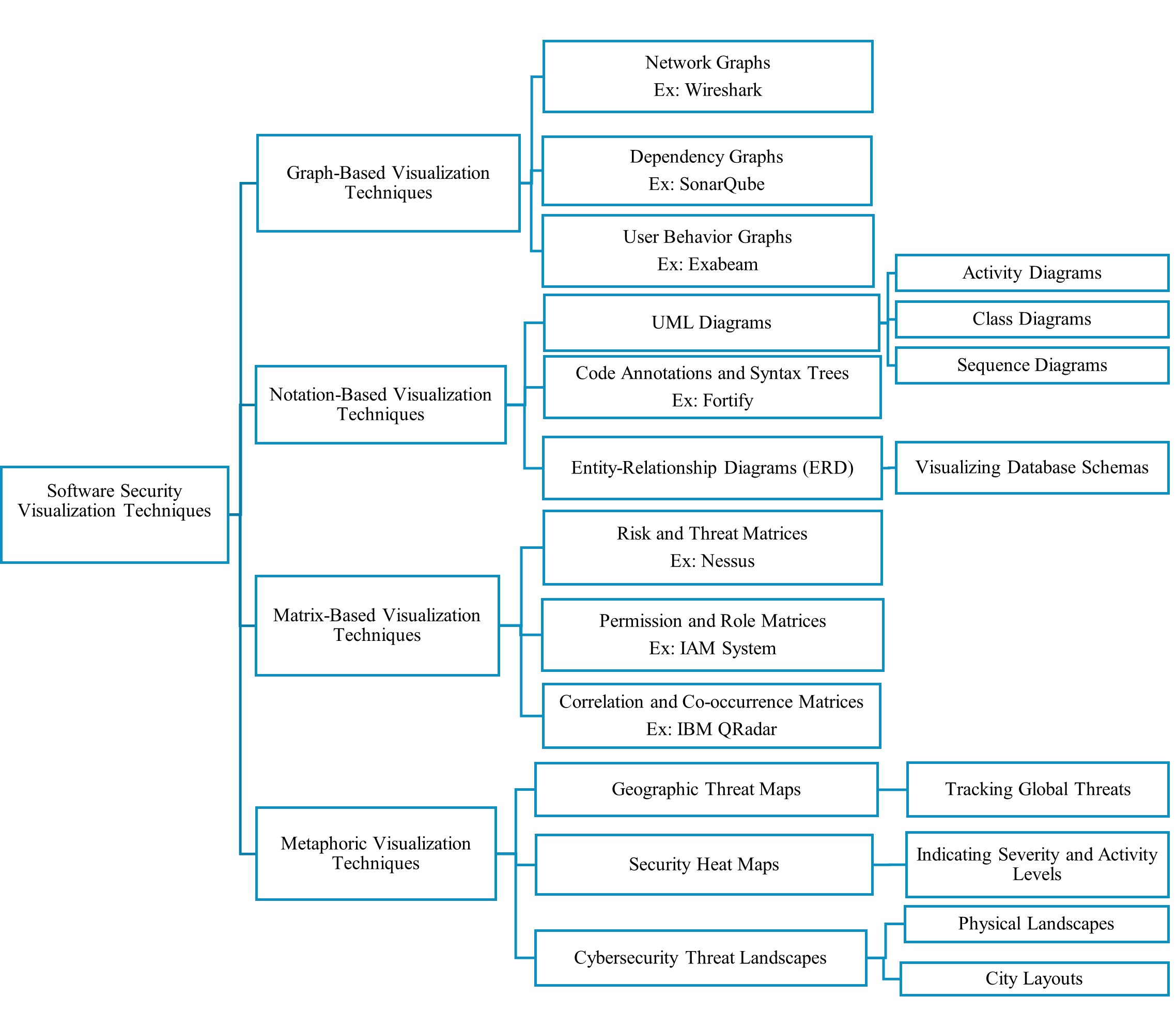}
 \caption{Software Security Visualization Techniques and Tools}
 \label{fig:Figure2}
\end{figure*}

A recent study has reviewed the importance of evaluating visualization tools to improve user performance and ensure reliable results. It has explored 3D visualization and software evolution techniques, highlighting their capacity to provide comprehensive views of software structures and changes over time \cite{korkut2023visualization,eberhard2023effects}. While 3D visualization offers significant advantages, its practical implementation in real-world software development environments is often limited by high computational costs and the steep learning curve required for effective use.

The visualization of code smells and software quality metrics was reviewed, emphasizing the role of identifying areas prone to security vulnerabilities \cite{dosReis2022code}. Integrating these visualizations into continuous integration/continuous deployment (CI/CD) pipelines is still a challenge, limiting their usefulness in agile development environments. Since software analysts need to read documents and source codes and synthesize trace data from multiple resources, the visualization for security analysis is usually labor-intensive \cite{aragones2023threat}. Recent studies highlighted that to resolve this problem, there is a requirement for automated analysis tools.

One such study proposed a threat-hunting architecture using a machine learning approach, which is a practical application of dynamic analysis in visualizing security structures and properties of software systems, providing real-time security assessments \cite{aragones2023threat}. While automation holds promise, its effectiveness in reducing human error and improving response times is still under debate, particularly in complex, rapidly evolving threat landscapes.

As shown in Figure \ref{fig:Figure2}, this review explores various software security visualization techniques essential for understanding and managing security risks in complex systems. These techniques are categorized into four main types: Graph-Based, Notation-Based, Matrix-Based, and Metaphoric Visualization. Each category uses different techniques and tools to represent security data, helping security professionals, developers, and non-technical users gain insights into system vulnerabilities, threats, and user behaviors.

\subsubsection{Graph-Based Visualization Technique} Graph-based visualization techniques focus on visualizing connections, data flows, and dependencies within software systems to identify relationships, detect anomalies, and monitor user behavior \cite{lagraa2024review}. Tools such as Wireshark are used for network graphs \cite{Arvind2023Network}, SonarQube for dependency graphs \cite{Biazotto2024Technical}, and Exabeam for user behavior graphs \cite{Sharma2024Developing}. These techniques offer real-time insights into network traffic, application dependencies, and user interactions, crucial for detecting security threats in large systems.

\subsubsection{Notation-Based Visualization Techniques} Notation-based visualization techniques visually represent the design and architecture of software systems, helping developers understand code structure and the flow \cite{wijesiriwardana2023software}, \cite{sinhabahu2020secure}. Techniques like UML diagrams (including activity, class, and sequence diagrams) \cite{mule2023implementation}, tools like Fortify for code annotations and syntax trees \cite{aravind2024codesheriff}, and Entity-Relationship Diagrams (ERD) are used to visualize database schemas \cite{fluchs2023evaluation}. This technique uses standardized diagrammatic representations to show the design and architecture of software systems and is useful in software design, debugging, and security risk assessment by mapping system components and their interactions.

\subsubsection{Matrix-Based Visualization Techniques} Matrix-based visualization techniques organize data into matrices to assess risks, permissions, and correlations within systems, offering a clear overview of user roles and access permissions \cite{wijesiriwardana2023software}. Tools such as Nessus for risk and threat matrices \cite{zhao2020network}, IAM Systems for permission, Role Matrices \cite{engstrom2023automated}, and IBM QRadar for Correlation and Co-occurrence Matrices to assess security risks and map user roles \cite{nour2023survey}. These metrics are valuable in enterprise security, ensuring compliance, and detecting unauthorized access.

\subsubsection{Metaphoric Visualization Techniques} Metaphoric visualization techniques use real-world metaphors, like geographic maps and city layouts, to intuitively represent complex security data \cite{moreno2024influence}. Geographic Threat Maps are used for tracking global threats \cite{Jiang2024Virtual}, Security Heat Maps for indicating severity and activity levels \cite{jiang2022systematic}, and Cybersecurity Threat Landscapes use physical and city layouts to represent security data \cite{rosado2022marisma,wijesiriwardana2023software}. These techniques help non-technical users understand security risks and threats, improving situational awareness and aiding in action prioritization.

\section{Software Security Visualization In Practice}

Software Security Visualization in Practice involves the application of visual techniques to enhance the understanding, monitoring, and management of security within software systems. As cybersecurity threats grow in complexity, traditional methods of analyzing security data often fall short. Visualization techniques bridge this gap by transforming complex data into intuitive, visual formats, enabling quicker and more accurate detection, analysis, and response to security threats. This practical approach to software security is critical for improving both the effectiveness and efficiency of cybersecurity efforts in real-world environments.

\subsection{Operational Security Visualization}

\begin{figure}[]
 \centering
\includegraphics[width=100mm]{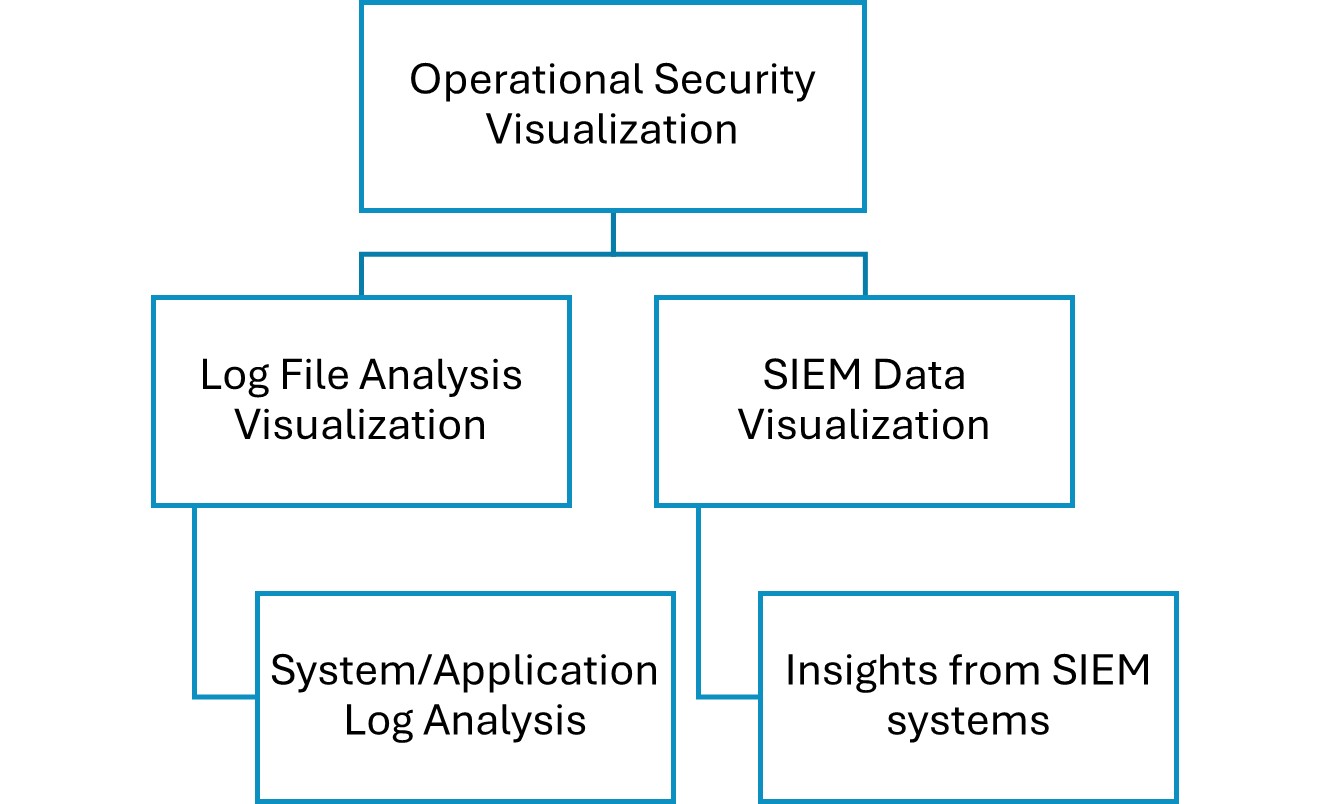}
 \caption{Operational Security Visualization}
 \label{fig:Figure3}
\end{figure}

Operational security visualization is a crucial aspect in the field of cybersecurity, and it focuses on the graphical representation of security-related data to analyze security incidents. This study involves the use of visualization techniques and tools to interpret complex datasets, identify vulnerabilities, and maintain the integrity of the system. Figure \ref{fig:Figure3} illustrates Operational Security Visualization, which includes techniques for analyzing log files and SIEM data to monitor and enhance security operations.

\subsubsection{Log File Analysis Visualization} Log file analysis visualization helps in identifying security vulnerabilities, understanding system behavior, and enhancing overall security posture. Recent studies have emphasized the importance of visualization techniques to manage and interpret network logs, categorizing various security visual analytics methods, and highlighting the challenges of extracting useful information from raw log data.

One study discussed how the process of visualizing cyber-physical systems involves transforming raw unstructured data into information that can be effectively represented and analyzed, underscoring the necessity of robust visualization tools in cybersecurity \cite{cobilean2023review}. However, this study did not extensively compare the effectiveness of different visualization tools, leaving a gap in understanding which tools are most effective for specific tasks. Another study by Korzeniowski and Goczyła highlighted recent advancements in automated log analysis, noting the benefits of standardizing log data formats to improve real-time monitoring, system health understanding, and root cause analysis \cite{korzeniowski2022landscape}. Despite these advancements, integrating standardized log formats into existing security infrastructures remains a challenge, particularly in large-scale systems. The role of log file analysis in cybersecurity assessments is emphasized, reviewing methodologies and tools for measuring and assessing enterprise security, identifying vulnerabilities, and improving overall security infrastructure \cite{aboelfotoh2019review}. While these methodologies are effective, their scalability and applicability across different types of enterprises are still underexplored, representing a significant research gap.

\subsubsection{SIEM Data Visualization} SIEM data visualization plays a crucial role in interpreting and managing the extensive amounts of security-related data collected by Security Information and Event Management (SIEM) systems. Recent research underscores the integration of big data analytics with SIEM to enhance the analysis and visualization of network security threats.

For instance, a study on effective security monitoring using efficient SIEM architecture emphasizes the use of advanced big data analytics tools to improve threat detection and visualization capabilities \cite{sheeraz2023siem}. Additionally, another study highlights the limitations of traditional SIEM tools, such as their struggle to handle large-scale data efficiently and the inability to detect sophisticated attacks. This research highlights the need for a holistic framework that incorporates advanced visualization techniques to boost threat detection and response, addressing the challenges of continuous monitoring and rapid analysis of substantial volumes of security data in modern enterprises \cite{jose2023framework}.

Another study focused on cyber situational awareness (CSA) visualizations, underscoring the need for sophisticated visualization techniques to support various stakeholders and improve threat understanding and response planning \cite{page2021prisma}. Despite these findings, the challenge of tailoring these visualizations to different user levels, particularly non-expert users, remains underexplored, presenting another gap in the current literature. Additionally, recent research has addressed the challenges of processing unstructured text logs in SIEM systems, emphasizing the importance of visualization in making sense of this data to detect patterns and anomalies \cite{reddy2023siem}.

These studies collectively demonstrate that effective visualization techniques are essential for improving the functionality and effectiveness of SIEM systems, enabling better threat detection, situational awareness, and overall cybersecurity. However, further research is needed to develop more advanced, user-friendly visualization tools that can handle the increasing complexity and volume of security data.

\subsection{Cyber Security Visualization}

Cybersecurity is a significant component of modern cybersecurity strategies, transforming complex data into visual formats that facilitate cyber threat understanding, detection, and mitigation. Figure 4 depicts Cybersecurity Visualization, which includes techniques for visualizing threat intelligence, phishing attacks, and malware behavior patterns to enhance security analysis and response.

\begin{figure}[H]
 \centering
\includegraphics[width=90mm]{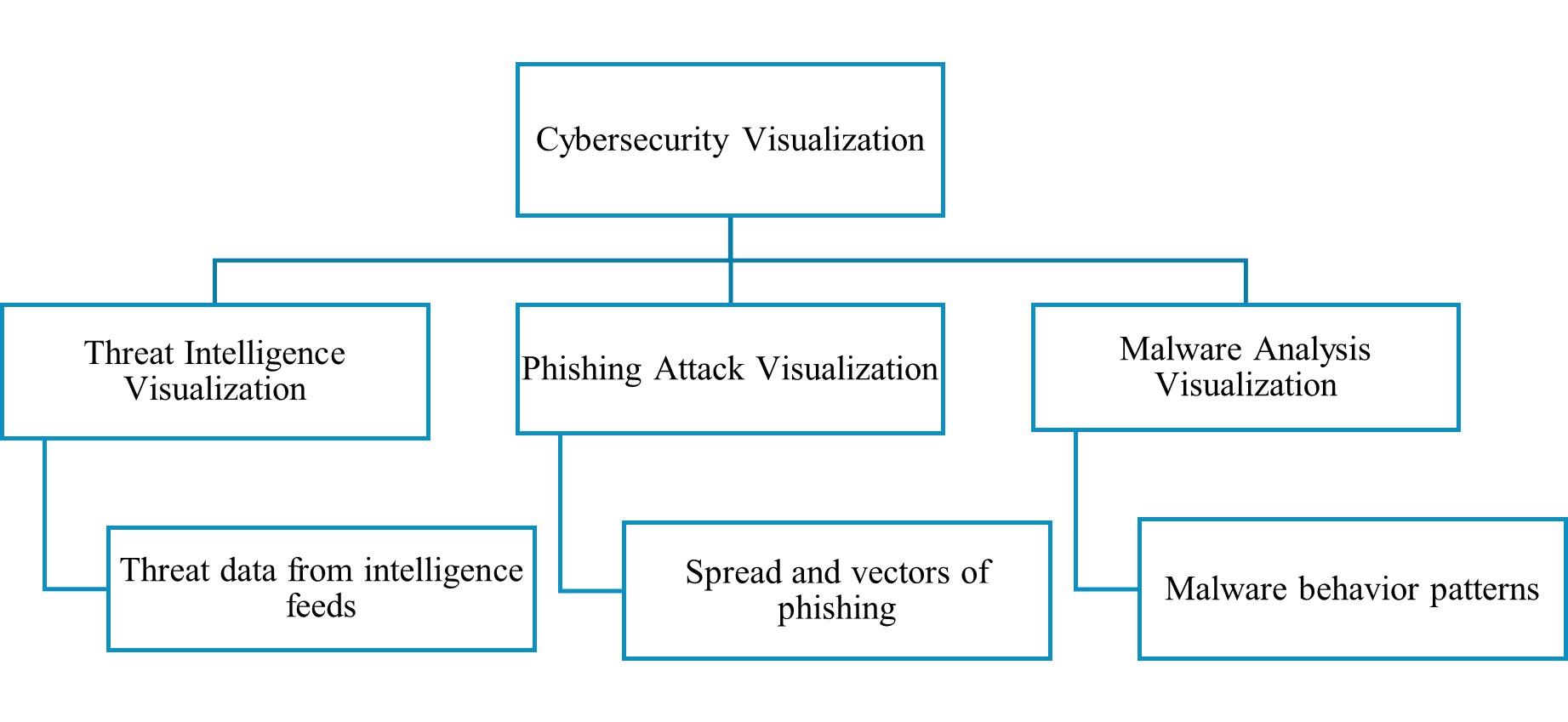}
 \caption{Cyber Security Visualization}
 \label{fig:Figure4}
\end{figure}

\subsubsection{Threat Intelligence Visualization} Threat intelligence visualization is vital for enhancing software security by transforming complex threat data into actionable insights. A recent study has emphasized the need for visualizing threat data from surface, deep, and dark web levels to aid in cybercrime detection and support law enforcement agencies \cite{Cascavilla2021Cybercrime}. Sun et al. highlighted the importance of real-time threat data visualization for proactive cybersecurity defense, providing a comprehensive review of recent research on cyber threat intelligence mining \cite{sun2023cyber}. While these studies underscore the importance of visualization, they often overlook the practical challenges of implementing these techniques in real-world scenarios, such as the need for real-time processing and integrating multiple data sources.

A study has discussed the use of natural language processing techniques to extract cyber threat intelligence from unstructured texts, demonstrating how visualization can help interpret large volumes of data and identify key indicators of compromise. It also highlighted the automation of CTI extraction and the role of visualization in supporting proactive decision-making and threat mitigation \cite{rahman2020cti}. However, the scalability of these techniques remains a concern, particularly in large organizations with vast amounts of unstructured data. Recent studies have demonstrated the need for standardized visualization techniques to improve the quality and sharing of technical threat intelligence, facilitating better collaboration and response to cyber threats \cite{sun2023cyber}. Despite these advances, the lack of universally accepted standards for threat intelligence visualization continues to hinder effective collaboration between organizations. Collectively, these studies underscore the critical role of visualization in understanding and responding to cyber threats, thereby enhancing the overall security posture of organizations.

\subsubsection{Phishing Attack Visualization} Phishing attack visualization is an essential tool in enhancing cybersecurity by transforming complex data into actionable insights to detect and mitigate phishing threats. Recent studies have underscored the significant role of software security visualization in the real-time detection and monitoring of phishing attacks. For instance, recent research highlights the importance of utilizing big data analytics within software security visualization to decrease the number of phishing attacks, emphasizing that visualizing large datasets can expose malicious patterns and enhance threat detection \cite{pejic2023phishing}. However, while these techniques are effective, their reliance on large data sets can also be a limitation, particularly in environments where data availability is restricted.

The effectiveness of machine learning-based visualization techniques in detecting phishing activities is highlighted, offering clear visual cues to identify suspicious behaviors \cite{alnemari2023phishing}. Patil and Arra focused on the role of visual tools in user awareness training, demonstrating how visualizing phishing threats can educate users and improve their response to such attacks \cite{patil2022phishing}. Despite these advancements, the challenge of integrating these tools into existing training programs and ensuring their accessibility to non-technical users remains underexplored. Another recent study provided a comprehensive review of cybersecurity measures, emphasizing the need for visualization in understanding and controlling phishing attacks \cite{verma2025cyber}. These studies collectively underscore the critical role of visualization in enhancing the detection, prevention, and response to phishing threats, thereby improving the overall security posture of organizations. However, further research is needed to improve the usability and accessibility of these tools across different organizational contexts.

\subsubsection{Malware Analysis Visualization} Malware analysis visualization is crucial in the fight against sophisticated cyber threats, offering a visual approach to understanding and mitigating malicious activities. A study highlighted the importance of Cyber Situational Awareness (CSA) in comprehending complex cyber threats, emphasizing the role of visualization in providing operational-level insights but noting gaps in visualizations for higher-level decision-making \cite{jiang2022systematic}. Another study has provided a systematic review of malware visualization techniques, showcasing their effectiveness in detecting, classifying, and identifying malware through static data visualization and network traffic monitoring \cite{alfagi2019malware}. Despite these findings, the study noted that the effectiveness of these techniques in large-scale, real-world environments still needs to be validated. Moawad et al. highlighted visualization-based malware detection, noting its efficiency and independence from domain experts, using image representation of malware binaries to enhance detection accuracy \cite{moawad2022malware}. The challenge of maintaining accuracy and reducing false positives in these visualizations remains a significant barrier to their broader adoption.

Recent studies have emphasized the importance of advanced visualization systems in malware analysis. One such study conducted a comprehensive survey on the integration of explainable AI (XAI) methods for malware hunting. This research evaluated various XAI techniques for their effectiveness in visualizing and analyzing malware \cite{saqib2024explainableai}. However, the challenge of integrating these advanced visualization techniques into existing malware analysis workflows remains a significant area for future research. These studies collectively underscore the significant role of visualization in malware analysis, enhancing the detection, classification, and mitigation of malware by transforming complex data into understandable visual formats, thereby supporting both operational and strategic cybersecurity efforts.

\section{Discussion}

This systematic review has provided an in-depth exploration of the current state of software security visualization techniques. Key findings indicate that visualization techniques are crucial in improving security monitoring, threat detection, and overall system understanding. The review categorized the main approaches into four types: graph-based, notation-based, matrix-based, and metaphor-based visualizations.

Graph-based visualization techniques, such as network graphs (Wireshark) and user behavior graphs (Exabeam), are primarily effective in depicting connections, data flows, and system dependencies. These visualizations are particularly useful for real-time monitoring of dynamic systems, helping detect anomalies in user behavior and network traffic. However, they face challenges in scalability and real-time integration, particularly in large, complex systems. Notation-based visualization methods, like UML diagrams and entity-relationship diagrams (ERD), provide clarity on system design and architecture. They are instrumental in mapping out system components and flows, which is essential for security assessments. However, these tools may require significant effort from developers and may not adequately capture dynamic security events, limiting their effectiveness in real-time threat detection. Matrix-based visualization, used by tools like Nessus and IBM QRadar, focuses on assessing risks, permissions, and correlations within a system. These visualizations provide a structured view of user roles, permissions, and identified threats, making them effective for managing large-scale enterprise security environments. However, their static nature and reliance on predefined matrices can limit their adaptability to real-time threats, especially in rapidly evolving attack scenarios. Metaphor-based visualization techniques, such as security heat maps and geographic threat maps, leverage real-world metaphors to represent complex security data. These methods are especially useful in communicating security risks to non-technical stakeholders, improving overall situational awareness. However, their reliance on simplified representations may obscure the complexities of security systems, leading to oversights in more nuanced threat environments.

\subsection{Recent Approaches}

Recent research has introduced several innovative approaches to enhance software security visualization. One of the emerging trends is the integration of big data analytics and machine learning with visualization techniques. This allows for more dynamic threat detection, particularly in Security Information and Event Management (SIEM) systems. SIEM platforms, when combined with advanced visualization tools, can handle larger volumes of data more efficiently, providing real-time insights into complex security landscapes. Furthermore, the application of deep learning to automate vulnerability detection in static and dynamic analysis environments has gained attention. By applying machine learning algorithms to large datasets, these approaches can enhance the accuracy of vulnerability detection, though challenges remain in reducing false positives and improving the interpretability of results. In addition to machine learning-driven approaches, recent research has explored adaptive visualization mechanisms such as Level of Detail (LoD) techniques integrated with city metaphors to dynamically adjust visual complexity based on user roles, tasks, and real-time security data, thereby reducing cognitive overload and improving vulnerability detection \cite{devendra2024lod}.

\subsection{Hybrid Visualization Techniques}

An emerging concept in the literature is using hybrid visualization techniques, combining the strengths of multiple approaches to provide a more comprehensive view of security risks. For example, integrating graph-based visualizations with matrix-based approaches enables a holistic view of system structure and risk correlations. This hybrid approach provides deeper insights into security vulnerabilities, offering real-time mapping of user roles, permissions, and system behavior. However, the practical implementation of hybrid techniques is still under research, with scalability and performance in large-scale environments being key challenges.

\subsection{Challenges}

While current software security visualization tools offer significant benefits, several ongoing challenges persist. A major concern is the need for scalable solutions capable of processing large datasets in real-time, as modern software systems generate vast amounts of data that require real-time analysis \cite{cui2022empirical}. Additionally, improving user accessibility is crucial. Many current tools are designed for experts, making them less intuitive for non-expert users, which hinders broader adoption \cite{braz2022less,mollerstrom2023engineering}. Enhancing user-centric design could make these tools more accessible and effective for a wider audience.

Another challenge is the lack of standardized metrics. The absence of consistent methods for assessing visualization performance makes it difficult to compare tools and measure their effectiveness across different studies \cite{mollerstrom2023engineering}. Establishing standardized benchmarks could help address this issue and improve the reliability of security visualization tools.

\section{Conclusion}

This review highlights the importance of visualization techniques in understanding, analyzing, and mitigating security threats by transforming complex data into easily interpretable formats. Dynamic analysis visualization supports detecting vulnerabilities in real time and also provides clear visuals that improve Cyber Situational Awareness (CSA) for everyone, especially non-experts, by focusing on threat information. However, several key challenges persist in the field. These include the need for scalable tools capable of processing large datasets in real-time as software systems grow in complexity, the lack of user-centric designs that hinder the accessibility of these tools to non-technical users, and the absence of standardized metrics to assess the effectiveness of visualization tools. These advancements will lead to more reliable and accessible software security solutions, ultimately enhancing the ability to defend against emerging security threats. 

Future research should focus on integrating machine learning, Artificial Intelligence, and real-time data analysis capabilities while designing comprehensive platforms that unify various visualization techniques to enhance software security. 

\bibliographystyle{IEEEtran}
\bibliography{references}

\end{document}